\documentclass[12pt,preprint]{aastex}

\def\lapprox{$_<\atop{^\sim}$}

\slugcomment{To appear in ApJ}

\shorttitle{Dust clearing}
\shortauthors{Brown et al.}

\begin{document}


\title{LkH$\alpha$ 330: Evidence for dust clearing through resolved
submillimeter imaging}


\author{{J.M. Brown\altaffilmark{1}, G.A. Blake\altaffilmark{2}, C. Qi\altaffilmark{3}, C.P. Dullemond\altaffilmark{4}}, D.J. Wilner\altaffilmark{3}}
\altaffiltext{1}{Division of Physics, Mathematics \& Astronomy, MC
105-24, California
Institute of Technology, Pasadena, CA 91125; jmb@astro.caltech.edu}
\altaffiltext{2}{Division of Geological \& Planetary Sciences, California
Institute of Technology, Pasadena, CA 91125}
\altaffiltext{3}{Harvard-Smithsonian Center for Astrophysics, 60 Garden Street,
Mail Stop 42, Cambridge, MA 02138}
\altaffiltext{4}{Max-Planck-Institut fur Astronomie, Koenigstuhl 17,
69117 Heidelberg, Germany}



\begin{abstract}
Mid-infrared spectrophotometric observations have revealed a small
sub-class of circumstellar disks with spectral energy distributions
(SEDs) suggestive of large inner gaps with low dust content. However,
such data provide only an indirect and model dependent method of
finding central holes.  We present here the direct characterization of
a 40 AU radius inner gap in the disk around LkH$\alpha$ 330 through
340 GHz (880 $\mu$m) dust continuum imaging with the Submillimeter
Array (SMA). This large gap is fully resolved by the SMA observations
and mostly empty of dust with less than 1.3 $\times$ 10$^{-6}$
M$_\odot$ of solid particles inside of 40 AU. Gas (as traced by
accretion markers and CO M-band emission) is still present in the
inner disk and the outer edge of the gap rises steeply -- features in
better agreement with the underlying cause being gravitational
perturbation than a more gradual process such as grain growth.
Importantly, the good agreement of the spatially resolved data and
spectrophometry-based model lends confidence to current
interpretations of SEDs with significant dust emission deficits as
arising from disks with inner gaps or holes. Further SED-based
searches can therefore be expected to yield numerous additional
candidates that can be examined at high spatial resolution.

\end{abstract}




\section{Introduction}
The conservation of angular momentum during cloud collapse leads to
the creation of circumstellar disks around young stars that provide an
environment for planetary system formation. These disks are widely
studied through their dust content that causes excess emission above
that expected from the stellar photosphere at near-IR to
mm-wavelengths. The physical conditions in circumstellar disks are
comparable to those inferred for the early stages of our own solar
system, with sizes of tens to hundreds of astronomical units (AU;
i.e. the distance between the Sun and the Earth) and masses of
{\lapprox}0.1$M_\odot$ (\citealt{beckwith96}, \citealt{mundy00}).  At
the other end of the expected evolutionary sequence are less massive
debris disks, which are older and exhibit weaker infrared (IR) excess
emission that is thought to arise from collisions of planetesimals
after the first generation of dust has coagulated or dissipated
(\citealt{aumann84}, \citealt{rieke05}).

Less is understood about how disks transition between these phases, an
interval over which significant evolution of proto-planetary bodies is
expected to occur. Thus, knowledge of how disks dissipate is vital for
our understanding of planetary system formation. Studies have shown
that the transition period lasts at most $\sim$3-5 Myr
(e.g. \citealt{strom89}; \citealt{beckwith90}), making the
identification of a sample of this important, short-lived phase
difficult. One indicator of intermediate systems is the presence of an
inner hole or gap indicating that the inner disk has evolved while the
outer disk has not. The presence of gaps may indicate that planets
have already formed in the disks and cleared the material around their
orbits, as has been shown in the case of TW Hya
\citep{setiawan07}. Such systems can therefore strongly constrain
models of planet formation, especially the role of gap formation and
disk-planet interactions in various planet migration scenarios that
lead to the creation of the ``hot Jupiters'' found to orbit much older
systems \citep{marcy05}. It is therefore essential to search for
further examples that either support or reject the gap hypothesis.

In theory, an inner gap in a proto-planetary disk can be identified
spectrophotometrically since hotter dust emits at shorter wavelengths
and dust temperature is primarily a function of radius in disks. Thus,
a gap causes a reduction in flux at specific wavelengths which can
then be related to radius. For gaps close to the star, the SED is
depressed over wavelengths of 1-15 \micron\, as the absence of hot
dust results in the infrared flux coming largely from the stellar photosphere,
rather than disk surface emission. To date, such emission ``deficits''
are the tool most widely used to infer the presence of gaps
(\citealt{calvet02}, \citealt{forrest04}), but spectrophotometric
signatures are indirect and notoriously difficult to interpret as
multiple physical scenarios can result in the same SED. For instance,
both a gap from a companion and large grain growth could lead to
missing mid-IR emission in the SED. Additional constraints and
detailed models are therefore required to distinguish between the
possible physical scenarios that are consistent with the observed
fluxes.

Disks emit strongly at submillimeter wavelengths where
interferometers like the Submillimeter Array (SMA) allow small
scale imaging.  Here we present some of the first direct evidence for
large inner holes in the form of a SMA 880 $\mu$m continuum map resolving
the 40 AU radius inner hole in the disk around LkH$\alpha$ 330, a young G3
pre-main sequence star in the Perseus star forming region.  Although the
distance to the Perseus molecular cloud is uncertain, here we adopt
250 pc in agreement with \citet{enoch06}. LkH$\alpha$ 330 has a mass of
2.5 M$_\odot$, an effective temperature of 5800 K and an age of $\sim$3 Myr
(\citealt{ob95}, \citealt{ck79}). We first present the relevant
observational details before turning to a discussion of the results and
their implications for SED-driven searches for gaps in circumstellar disks.

\section{Observations}
Mid-IR spectrophotometry of LkH$\alpha$ 330 was acquired as part of
the Spitzer ``From Cores to Disks'' (c2d) Legacy Science project. The
resulting SED is shown in Figure \ref{fig:sed}. Out of a sample of over
100 spectra in the c2d first look program, LkH$\alpha$ 330 was one of
only 5 disks which showed SED features characteristic of an inner hole
\citep{brown07}. Those disks visible from the northern hemisphere have
been targeted for high spatial resolution follow-up imaging at IR through
mm-wavelengths.

Dust emission measurements of LkH$\alpha$ 330 were acquired with the
Submillimeter Array (SMA)\footnote{The Submillimeter Array is a joint
project between the Smithsonian Astrophysical Observatory and the
Academia Sinica Institute of Astronomy and Astrophysics, and is
funded by the Smithsonian Institution and the Academia Sinica.}
(\citealp{ho_m04})
 on 2006 November 11 and 19 using the very extended
configuration of seven of the 6 meter diameter antennas, which provided
baselines ranging in length from 80 to 590 meters. Double sideband (DSB)
receivers tuned to 341.165 GHz provided 2 GHz of bandwith/sideband,
centered at an Intermediate Frequency (IF) of 5 GHz. The DSB system
temperatures ranged from 210 to 870 K, and LkH$\alpha$ 330 was observed
from hr angles of -4.5 to 4.5 to yield a synthesized beam of
0\farcs28x0\farcs33. Calibration of the visibility phases and amplitudes
was achieved with observations of 3C111, typically at intervals of 25 minutes.
Measurements of Uranus and Titan provided the absolute scale for the
flux density calibration and the uncertainties in the flux scale are
estimated to be 15\%. The data were calibrated using the MIR software
package ({\tt http://cfa-www.harvard.edu/$\sim$cqi/mircook.html}), and
processed with Miriad \citep{sault95}.

\section{Results}
\begin{figure}
\includegraphics[angle=90, scale=0.375]{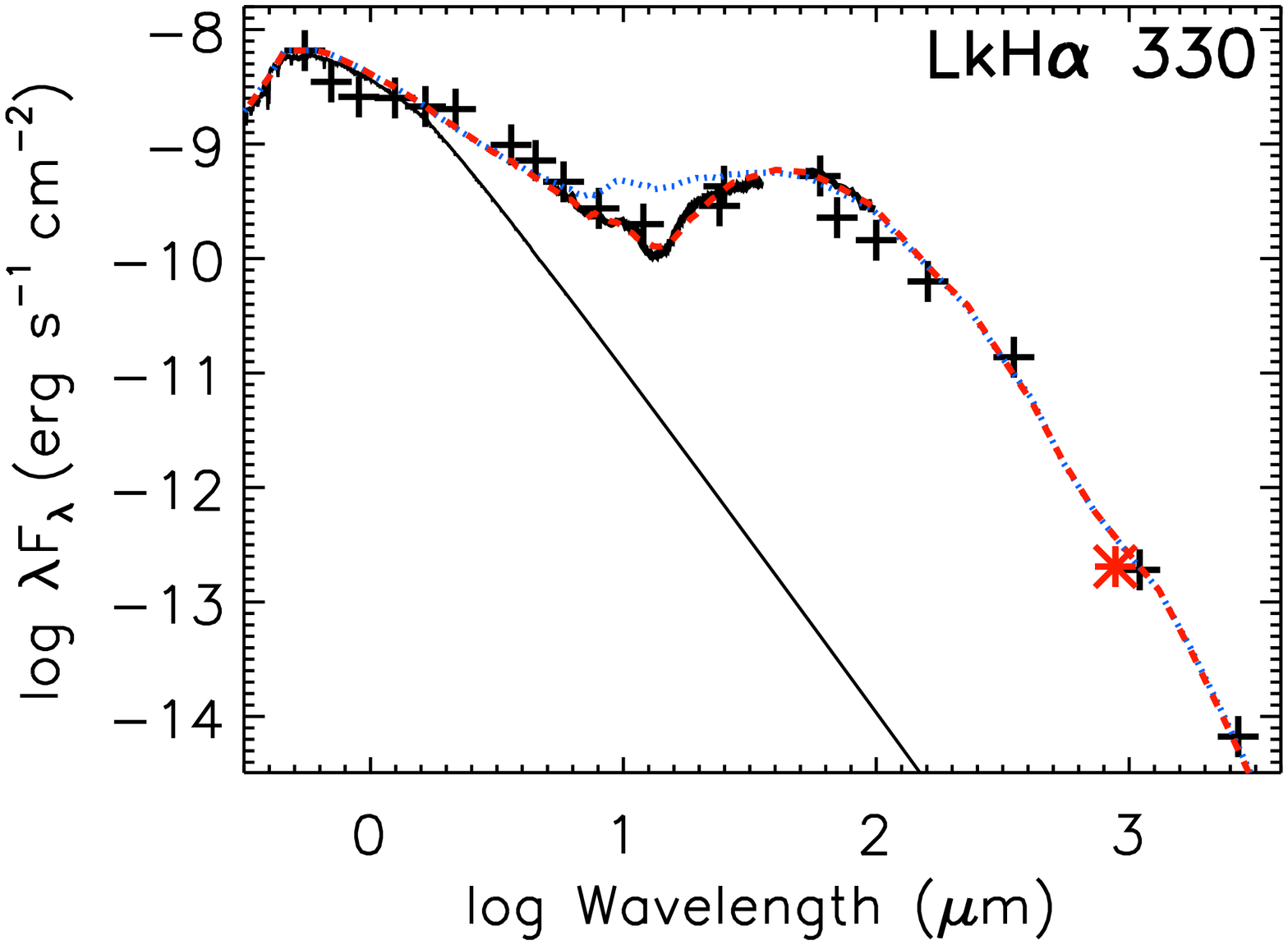}
\caption{The SED of LkH$\alpha$ 330.  The dashed curve depicts a
  disk model with a hole of radius 40 AU, while the dotted line is
  the equivalent model with no hole. The solid black curve is the
  stellar photosphere. The star is the total flux in the SMA image
  and is consistent with previous photometry.
\label{fig:sed}}
\end{figure}

Fits to the optical through millimeter-wave SED yield an estimated gap
outer radius of 40 AU (Fig. \ref{fig:sed}, \citealp{brown07}). With a
sufficiently sharp transition, aperture synthesis observations should
detect a null in the flux versus ($u,v$)-distance as opposed to the
smooth drop off in flux associated with power-law mass surface density
profiles that characterizes most classical T Tauri star disks
\citep{aw07}. As Figure 2 shows, such a null is indeed present in
the SMA data at a distance of $\sim$700-800 nsec (or 210-240
k$\lambda$ at 340 GHz).

When Fourier transformed to the image plane, the SMA observations
clearly resolve the size, orientation and radial structure in the
LkH$\alpha$ 330 disk (see Figure \ref{fig:model}, top panel). The hole
is approximately the size of the synthesized beam at 0\farcs33,
corresponding to a hole radius of 41 AU. The integrated flux of the
disk is 60 mJy, and the 1-sigma Root Mean Square (1$\sigma$ RMS) noise
in the map is 2.3 mJy. The outer disk has a radius of 0\farcs5, or 125
AU at $d$=250 pc; and the position angle of the disk major axis is
75$^{\circ}$. For a disk with azimuthal symmetry the inclination angle
is 40$^{\circ}$, and this inclination causes the most prominent
asymmetries where more flux is seen east-west than north-south. The
edges of the disk where more dust lies along the line-of-sight have
higher column densities, creating two bright regions aligned with the
major axis of the disk. However, other significant asymmetries remain
in the data.

The hole is largely empty of dust and the significant intensity
contrast between the inner and outer disk indicates a large mass
surface density contrast -- even for mm-sized grains. The flux within
the hole is below the 1$\sigma$ RMS of 2.3 mJy. Using the flux
to disk mass conversion of \citet{beckwith90}, this places a limit on
the amount of mass in the hole of only 1.3 $\times$ 10$^{-4}$
M$_\odot$ for a gas:dust ratio of 100. The boundary between hole
and outer disk is abrupt with the flux dropping 25\% in less
than 10 AU and 50\% in less than 20 AU.

The 2-D radiative transfer code RADMC \citep{dd04} was used to
simultaneously model both the resolved image and the SED. This model
assumes a passive disk, which merely reprocesses the stellar radiation
field. In order to fit the missing dust emission, the code was
adapted to reduce the dust density over a specific region to create a
gap in the disk. The resulting image was resampled in Miriad using the
same ($u,v$)-plane sampling as the SMA data so the two are directly
comparable (see Figure \ref{fig:model}, middle panel).

In the model, the disk is assumed to be flared such that the surface
height, $H$, varies with radius, R, as $H/R \propto R^{2/7}$
\citep{cg97}. The models do not calculate the hydrostatic equilibrium
self-consistently but physicality was checked with an a posteriori
calculation at the outer disk edge. Thus, the pressure scale height
(H/R) is anchored at 0.17 at the outer disk edge, in this case, 300
AU.  The dust composition is set to a silicate:carbon ratio of 4:1,
although moderate changes in this ratio have little effect at 880
\micron.  Only amorphous silicate is included as no crystalline
features are seen in the Spitzer spectrum. The grain sizes range from
0.01 $\mu$m to 10 cm with a power-law index of -3.5 \citep{draine06},
constrained by both the weak silicate features and the well-measured
Rayleigh-Jeans slope. The total disk mass, assuming a gas of solar
composition, is 0.017 M$_\odot$ \citep{ob95}. The disk inner edge was
set at 0.26 AU -- the radius at which the dust sublimation temperature
of 1500 K is reached.

The gap is represented in the model by three parameters: both an inner
and an outer gap radius and a density reduction factor. The best fit
model was found by performing a $\chi^2$ minimization on the inner and
outer gap radii to find the best fit to the SED (see \citealt{brown07}
for further details). This model was then compared to the image to
assess the validity of the SED interpretation. The SED best fit model
matched the image remarkably well, lending confidence to current
spectrophotometry-based models of transitional disks. Both the
submillimeter image and the SED indicate that the gap has an outer
radius of $\sim$40 AU. The SMA image places no constraint on the inner
gap radius but the SED requires 0.2 M$_{\rm lunar}$ of dust closer
than 1 AU to the star to account for the near-IR excess observed at
wavelengths greater than $\sim$2 $\mu$m (see Fig. \ref{fig:sed}).
Within our solar system, this gap corresponds to the area between
Earth's orbit and the inner Kuiper Belt.

The density reduction factor is severely constrained by the low flux
at large ($u,v$) distances in the SMA data (or R$<$40 AU in the SMA
image). The best fit model consistent with the SED has a density
reduction of 1000 within the gap, corresponding to 3.1 $\times$
10$^{-5}$ M$_\odot$ of material as compared to the $<$1.3 $\times$
10$^{-4}$ M$_\odot$ upper limit derived from the image. For
comparison, the model without the gap contains some 0.01 M$_\odot$ of
material within the same region. All of these models assume a gas:dust
ratio of 100, and still require a ring of dust within an AU of the
star to fit the observed near-IR excess. However, the required mass
within 1 AU is well below the detection limit of the SMA image.

In order to further investigate the abruptness of the transition
between hole and outer disk, a gap edge was introduced into the RADMC
model such that the dust mass surface density rises logarithmically
over a range of radii around the outer gap radius as one moves
outward. This modeling confirms that the aperture synthesis data are
consistent with a step function. Numerically, the transition region
between the hole and the outer disk can be no larger than 5 AU given
the present spatial resolution and dynamic range of the SMA data.
Such a steep transition is more consistent with truncations induced by
gravitational instabilities than a more gradual process such as dust
settling and coagulation (\citealt{crida06}, \citealt{dd05},
\citealt{tanaka05}).

\section{Discussion}
As the bottom panel of Fig. 3 shows, significant asymmetries remain
in the data which cannot be explained by an axisymmetric disk.
Excess flux can be seen in the right side of the disk and a
deficit of flux can be seen in the north. Such asymmetries would be
expected from gravitational perturbation caused by a large planet or
binary companion.  For planetary companions of sufficient mass, some
disk gas is expected to be transported across the gap while the dust
transport is strongly inhibited \citep{alexander07}. In order
to verify or refute the presence of close companions, high dynamic
range diffraction-limited infrared searches with 8-10m class telescopes
will be necessary.

The presence of material close to the star, making the density
reduction a gap rather than a hole, places constraints on many
physical processes which might produce the gap. While the only
indication of dust close to the star is the 1-10 \micron\, excess, we
have two additional indications that there is gas close to the
star. First, LkH$\alpha$ 330 displays H Balmer $\alpha$ emission, with
equivalent width measurements ranging from 11 to 20 \AA\,
(\citealt{fernandez95}, \citealt{ck79}), and so is still accreting
gas.  Second, emission from warm (850 K) molecular gas is seen in the
4.7 \micron\, $v$=1$\rightarrow$0 rovibrational emission lines of CO
(Salyk et al., in prep.). This reservoir must be close to the star to
reach such temperatures. Interestingly, no CO emission is detected
from CoKu Tau 4, recently discovered to be a circumbinary disk with a
10 AU inner hole \citep{ireland08}.

One proposed process for quickly clearing the inner disk region is
photoevaporation (\citealt{clarke01}, \citealt{alexander06}). An inner
hole occurs when the photoevaporation rate driven by the ionizing flux
from the central star matches the viscous accretion rate. However,
this condition is only effectively fulfilled when accretion rates are
low and would result in no gas or dust close to the star for gap radii
of several tens of AU.  Photoevaporation is thus unlikely to be
responsible for the 40 AU radius gap observed in the LkH$\alpha$ 330
disk. In contrast to photoevaporation, a recently proposed MRI-driven
model predicts dust clearing, while substantial gas remains in the
inner disk and accretion rates are high \citep{chiang07}. Such a model
could account for the case of LkH$\alpha$ 330, but a close
examination of the accretion properties of the central star and the
disk viscosity is still needed.

An alternative explanation to the physical removal of the dust is that
it has grown beyond the size at which it efficiently radiates as a
blackbody so that it no longer emits strongly in the mid-IR and
submillimeter \citep{tanaka05}.  Within any realistic distribution of
dust grain sizes, even a minimum grain diameter of 50 \micron\,
significantly overproduces the flux in the 10 \micron\, region with no
density reduction. Thus, grain growth to very large sizes is needed
for this scenario. The sharp cutoff in dust mass surface density
between the inner and outer disk is also difficult to reconcile with
dust coagulation models given the long orbital timescales and smooth
variation in disk properties expected at 40 AU.

To summarize, LkH$\alpha$ 330 presents a dramatic case of a disk
evolving from the inside out rather than smoothly throughout the disk
as would be expected in alpha-viscosity models of disk evolution. The
40 AU radius gap is largely empty of dust but gas does remain. The
outer edge of the gap rises steeply, indicating an abrupt change in
disk properties and limiting plausible explanations as to the origin
of the gap. Importantly, the good agreement of both data and
model lends confidence to current interpretations of SEDs with
significant dust emission deficits as arising from disks with inner
holes. Further SED-based searches can therefore be expected to yield
numerous additional candidates that can be examined at high spatial
resolution. Ultimately, in such studies it will be critical to not
only image the dust but to provide estimates of the gas:dust ratios in
the outer and inner disk if the different possible gap creation
scenarios are to be disentangled. A combination of spatially resolved
imaging, ultimately with the Atacama Large Millimeter Array (ALMA), to
probe the dust and high resolution infrared through millimeter-wave
spectroscopy to trace the gas content will provide the critical
observational characterization needed for a quantitative understanding
of these interesting objects.

\acknowledgments {The authors wish to thank SMA personnel for their
assistance in acquiring the very extended configuration data reported here,
and the support provided by the NASA Spitzer Space Telescope Legacy Science
Program, through Contract Numbers 1224608 and 1230779 issued by the
Jet Propulsion Laboratory, California Institute of Technology under
NASA contract 1407, and the NASA Origins of Solar Systems Program
through Grants NNG05GH94G and NNG05GI81G.}

\bibliographystyle{apj}

\begin{figure}
\includegraphics[angle=90, scale=0.375]{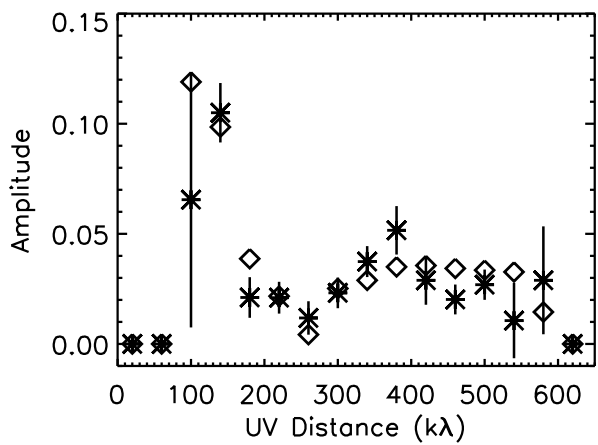}
\caption{SMA ($u,v$)-plane 340 GHz continuum data, with the
projected baseline length in kilolambda and the flux in Jy. Stars
denote the observations (with error bars), open diamonds the disk
with R=40 AU gap model sampled at the same antenna spacings.
\label{fig:uvdata}}
\end{figure}

\begin{figure}
\vskip 0.15in
\begin{minipage}{0.9\linewidth} 
\centering
\includegraphics[angle=-90, scale=0.28]{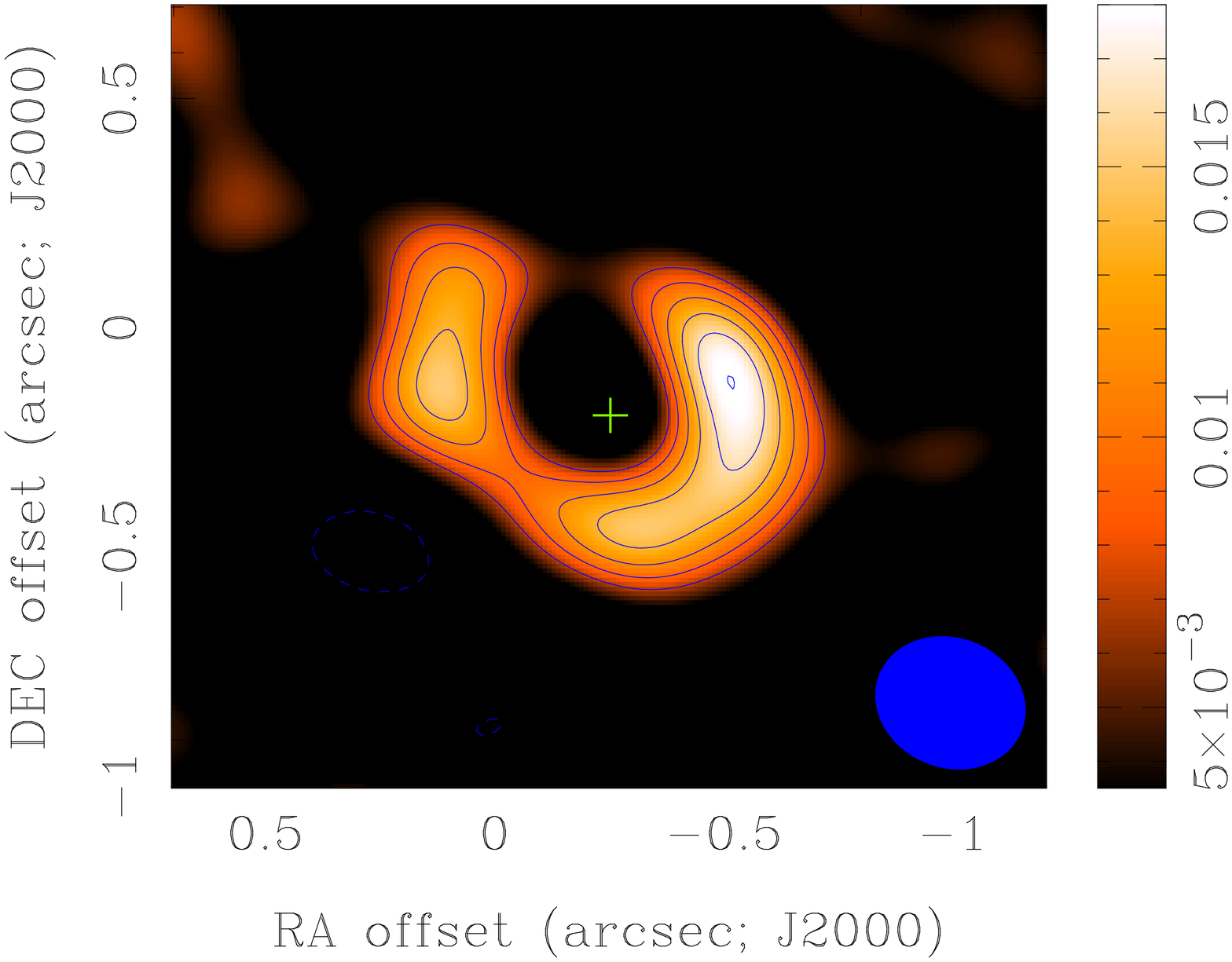}
\end{minipage}
\begin{minipage}{0.9\linewidth} 
\centering
\includegraphics[angle=-90, scale=0.28]{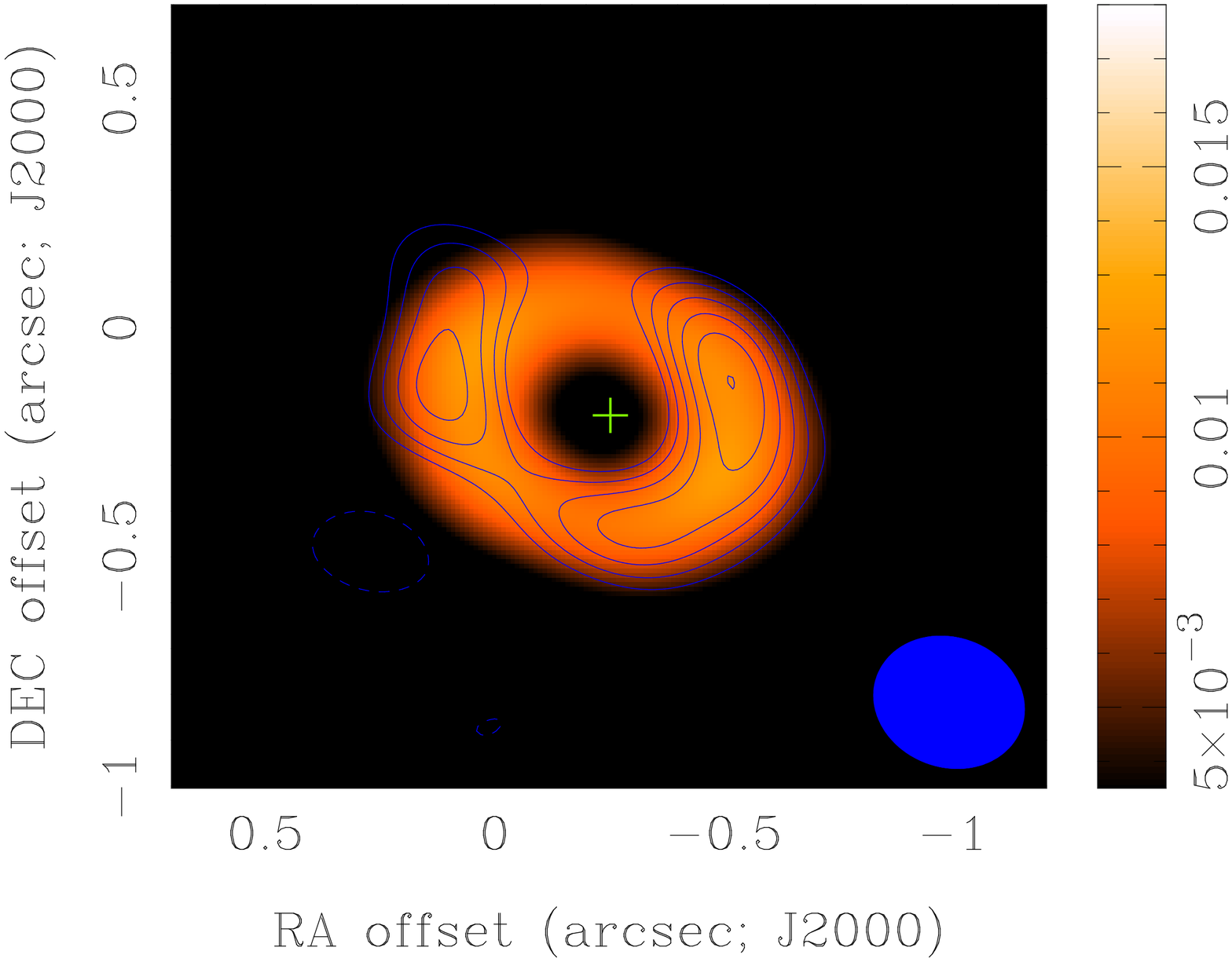}
\end{minipage}
\begin{minipage}{0.9\linewidth} 
\centering
\includegraphics[angle=-90, scale=0.28]{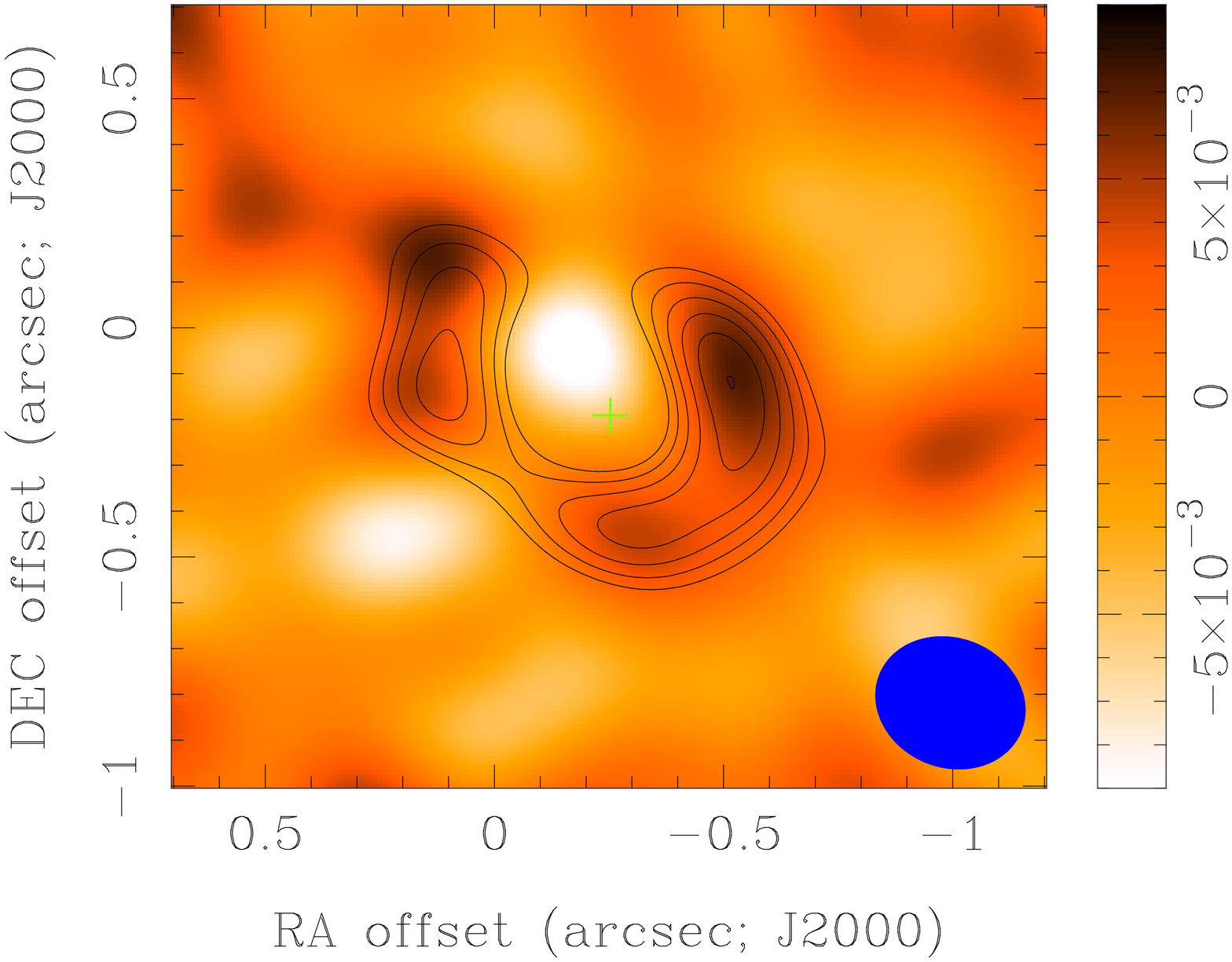}
\end{minipage}
\caption{(Top) The SMA 340 GHz dust continuum image of LkH$\alpha$ 330
clearly shows an inner disk hole of approximately 40 AU radius.
The 0\farcs28x0\farcs33 beam is plotted at bottom right.
(Middle) The model of LkH$\alpha$ 330 in color
overlaid with 1-sigma contours from the data, beginning at
3-sigma with the beam in the lower right corner. The model determines
that the hole has a radius of 40 AU. (Bottom)
The model subtracted from the data. Darker regions correspond to excess
flux in the data and lighter regions correspond to excess flux in the model.
The flux scale in the residuals plot is expanded by a factor of three
compared to the data plots. Significant asymmetries remain, particularly
in the north, which cannot be accounted for by an axisymmetric disk.
\label{fig:model} \vspace{0.6cm}}
\vskip 0.05in
\end{figure}

\end{document}